\documentclass[9pt,technote]{IEEEtran}

\usepackage{cite}      

\usepackage{graphicx}  

\usepackage{url}       

\usepackage{amsmath}   
\newtheorem{theorem}{\textbf{Theorem}}
\newtheorem{definition}{\textbf{Definition}}
\newtheorem{corollary}[theorem]{\textbf{Corollary}}

\begin{document}
%
\title{On the Unicity Distance of Stego Key }
%
%
\author{Weiming~Zhang, and~Shiqu~Li 
\thanks{Weiming Zhang and Shiqu Li are both with the Department of Applied Mathematics,
University of Information Engineering, P.O. Box 1001-747,
Zhengzhou 450002 P.R. China. Email: nlxd\_990@yahoo.com.cn}}
%
%
%
\markboth{Journal of \LaTeX\ Class Files,~Vol.~1, No.~11,~November~2002}{Shell \MakeLowercase{\textit{et al.}}: Bare Demo of IEEEtran.cls for Journals}
%



\maketitle

\begin{abstract}
Steganography is about how to send secret message covertly. And
the purpose of steganalysis is to not only detect the existence of
the hidden message but also extract it. So far there have been
many reliable detecting methods on various steganographic
algorithms, while there are few approaches that can extract the
hidden information. In this paper, the difficulty of extracting
hidden information, which is essentially a kind of privacy, is
analyzed with information-theoretic method in the terms of unicity
distance of steganographic key (abbreviated stego key). A lower
bound for the unicity distance is obtained, which shows the
relations between key rate, message rate, hiding capacity and
difficulty of extraction. Furthermore the extracting attack to
steganography is viewed as a special kind of cryptanalysis, and an
effective method on recovering the stego key of popular LSB
replacing steganography in spatial images is presented by
combining the detecting technique of steganalysis and correlation
attack of cryptanalysis together. The analysis for this method and
experimental results on steganographic software ``Hide and Seek
4.1" are both accordant with the information-theoretic conclusion.
\end{abstract}


%
\IEEEpeerreviewmaketitle

\section{Introduction}
%
%
%
%
\PARstart{S}{teganography} is an important branch of information
hiding, and it is about how to send secret message covertly. The
attacks to steganography (i.e. steganalysis) mainly include
passive attack, active attack, and extracting attack. A passive
attacker only wants to detect the existence of the embedded
message, while an active attacker wants to destroy it. The purpose
of an extracting attacker is to obtain the message embedded into
the innocent data. So there are three kinds of security for
different attacks respectively, i.e. detectability, robustness and
difficulty of extraction.

The theoretic study about steganography has always been concerning
the detectability, and there have been many literatures that model
the detectability with information-theoretic method or in the
terms of computational complexity
\cite{cac,zol:fed,kat:pet,hop:lan}. On the other hand, references
\cite{ett,som:mer,mou:sul} think of the information hiding problem
with active attackers as a ``capacity game'', and define the
robustness using the ``hiding capcity''. Although robustness is
mainly concerned in watermarking problem, it, as the measure of
efficiency, is also important for steganography. And references
\cite{vol:pun,cha:mem,mos:cha:new,zha:li} analyze the relation
between the detectability and robustness.

Similar with the theoretic field, the study about actual
steganalysis has also being centering on detecting technique. And
there have been many detecting methods for a variety of
steganographic algorithms such as \cite{wes:pfi,fri:gol,zha:pin}.
However, there are only a few papers about extracting attack.
Chandramouli \cite{cha} studies how to make extracting attack on
spread spectrum steganography for a special scenario in which the
same message is sent twice in the same image with different
strength factors. Fridrich et al. \cite{fri:gol:sou} show how to
get the hidden message through recovering the key of LSB
steganography on JPEG images such as ``F5 \cite{wes} and Outguess
\cite{pro}''. And recently in \cite{fri:gol:sou:hol} Fridrich et
al. extent their approach to spatial domain. Another extracting
approach to LSB steganography on JPEG images is presented by Ma et
al. \cite{ma:zha}.

The extracting attack on steganography can be viewed as a special
kind of cryptanalysis. In fact for most of steganographic systems
the message is required to be encrypted before it is hidden.
Therefore, when facing the model of "encrytion+hiding", a
cryptanalyst has to analyze a ``multiple cipher". Fridrich et al.
\cite{fri:gol:sou} analyze the complexity of searching stego-key:
If there is some recognizable structure in the steganographic
communication, one can use it as a sign to searching the key by
dictionary attack or brute-force search; otherwise, searching
process should try all encryption keys for every possible
stego-key, so the complexity of brute-force search becomes
proportional to the product of the number of stego and crypto
keys. That means that the extraction and decipher should be done
together. Obviously a cryptanalyst hope that the two tasks can be
finished independently. And the extracting attack just solve the
problem how to extract the embedded sequence without regard to
encryption algorithm.

In this paper, the difficulty of extraction, which is essentially
a kind of privacy, is studied with information-theoretic method in
the terms of unicity distance of stego key. Unicity distance is
just the minimum number of data needed by the attacker to recover
the stego key, which can exactly grasp the concept on ``difficulty
of extraction'' for key based stegonography. The relations between
key rate, message rate, hiding capacity and unicity distance are
analyzed. And it is proved that unicity distance is directly
proportional to the entropy of stego-key, and inversely
proportional to ``hiding redundancy" which is the difference
between the hiding capacity and message rate.

As mentioned above, our conclusion comes from the basic idea that
extracting attack on steganography is a special kind of
steganalysis. Therefore this problem can be solved by combining
traditional techniques of cryptanalysis and steganalysis together.
As an example, we present an extracting approach on random LSB
replacing steganography of spatial images, which is based on some
detecting techniques in steganalysis and the idea of correlation
attack \cite{sie} in cryptanalysis. One contribution of our attack
is that it can accurately estimate the amount of necessary data.
With this method, we make a successful extracting attack on
steganographic software ``Hide and Seek 4.1" \cite{mor} which is
found in the United States recently \cite{spy}. Experimental
results on ``Hide and Seek 4.1" are accordant with the analysis
for our extracting algorithm, which also verify the validity of
the information-theoretic conclusion.

The rest of this paper is organized as follows. The main theorem
on unicity distance of stego key is given in Sect. II. And in
Sect. III a method of recovering stego key -- ``correlation
attack" -- on LSB replacing steganography of spatial images is
presented. The experimental results on attacking ``Hide and Seek
4.1" is given in Sect. IV. And the paper concludes with a
discussion in Sect. V.

\section{Information-Theoretic Analysis for the Unicity Distance of Stego
key}

\subsection{Notations and Definitions}

For the information-theoretic analysis, we use the following
notations. Random variables are denoted by capital letters (e.g.
$X$), and their realizations by respective lower case letters
(e.g. $x$). The domains over that random variables are defined are
denoted by script letters (e.g. $\mathcal{X}$). Sequences of $N$
random variables are denoted with a superscript (e.g.
$X^{N}=(X_{1},X_{2},\cdots ,X_{N})$ which takes its values on the
product set $\mathcal{X}^{N}$). And we denote entropy and
conditional entropy with $H( \cdot )$ and $H(\cdot|\cdot)$
respectively.

A general model of a stegosystem can be described as follows. The
embedded data $M$ is hidden in an innocuous data $\widetilde{X}$,
usually named cover object, in the control of a secret stego key
$K$, producing the stego object $X$. The stego key is shared
between the sender and receiver but is secret for the third party.
And the receiver can extract $M$ from $X$ with the stego key $K$.
An extracting attacker wants to recover the embedded message or
the stego key through the stego object (Maybe he can use some side
information, for example part knowledge about the cover object).

Assume that the cover object data is a sequence $\widetilde{X}^N =
(\widetilde{X}_1 ,\widetilde{X}_2 , \cdots ,\widetilde{X}_N )$ of
independent and identically distributed (i.i.d) samples from
$P(\widetilde{x})$. Because the embedded message usually is cipher
text, we assume that it is a sequence $M^N  = (M_1 ,M_2 , \cdots
,M_N )$ of independent and uniformly distributed, and independent
of $\widetilde{X}^N$. The stego key $K$ is independent of the
message and cover object.

Now we describe a formal definition of steganographic code which
is introduced by Moulin et al. \cite{mou:sul,mou:wan}. First of
all, the embedding algorithm of a stegosystem should keep
transparency that can be guaranteed by some distortion constraint.
A distortion function is a nonnegative function $d:{\cal X} \times
{\cal X} \to {\cal R}^ + \cup \{0\}$, which can be extended to one
on N-tuples by $d(x^N ,y^N ) = \frac{1}{N}\sum\limits_{i = 1}^N
{d(x_i ,y_i )}$.

\bigskip
\begin{definition}
$^{[7]}$ A length-$N$ steganographic code subject to distortion
$D$ is a triple $({\cal M},f_N ,\phi _N )$, where
\begin{itemize}
\item ${\cal M}$ is the message set of cardinality $|{\cal M}|$;

\item $f_N :{\cal X}^N  \times {\cal M} \times {\cal K} \to {\cal
X}^N$ is the embedding algorithm mapping a sequence
$\widetilde{x}^N$, a message $m$ and a key $k$ to a sequence $x^N
= f_N (\widetilde{x}^N ,m,k)$. This mapping is subject to the
distortion constraint
\begin{eqnarray*}
\sum\limits_{\widetilde{x}^N  \in {\cal X}^N } {\sum\limits_{k \in
{\cal K}} {\sum\limits_{m \in {\cal M}}{\frac{1}{{|{\cal M}|
\cdot |{\cal K}|}}P(\widetilde{x}^N )} } }\\
{\cdot}\: d(\widetilde{x}^N ,f_N(\widetilde{x}^N ,m,k)) \le D
\enspace ;
\end{eqnarray*}

\item $\phi _N :{\cal X}^N  \times {\cal K}^N  \to {\cal M}$ is
the extracting algorithm mapping the received sequence $x^N$ with
the key $k$ to a decoded message $\widehat{m} = \phi_N (x^N ,k)$.
\end{itemize}
\end{definition}
\bigskip

A cover channel is a conditional $p.m.f.$ (probability mass
function) $q(x|\widetilde{x}):{\cal X} \to {\cal X}$. The compound
cover channel subject to distortion $D$ is the set
\[
Q= \{ q(x|\widetilde{x}):\sum\limits_{\widetilde{x},x}
{d(\widetilde{x},x)q(x|\widetilde{x})P(x)} \le D\}\enspace .
\]
The length-$N$ memoryless extension of the channel is the
conditional $p.m.f.$
\[
q(x^N |\widetilde{x}^N ) = \prod\limits_{i = 1}^N {q(x_i
|\widetilde{x}_i )}, \,\, \forall N \ge 1 \enspace .
\]
For a length-$N$ steganographic code, define the message rate and
key rate as
\[R_m  = \frac{{H(M)}}{N},\enspace R_k =\frac{{H(K)}}{N}\]
respectively. And define the probability of error as
$P_{eN}=P(\phi _N (X^N ,K) \ne M)$. The hiding capacity is the
supremum of all achieve message rates of steganographic codes
subject to distortion $D$ under the condition of zero probability
of error (i.e. $P_{e,N} \to 0 \mbox{ as } N \to \infty $).

Because we disregard the active attacker and assume that $K$ is
independent of $M$ and $\widetilde{X}$, the results of
\cite{mou:sul,mou:wan} imply that the expression of hiding
capacity for steganographic code can be given by
\begin{equation}\label{eqn1}
C(D) = \mathop{\max }\limits_{q(x|\widetilde{x}) \in Q}
H(X|\widetilde{X})\enspace .
\end{equation}

Because $C(D)$ is the maximum of the conditional entropy through
all cover channels subject to $D$ distortion, $C(D)$ just reflects
the hiding ability of the cover-object within the distortion
constraint. So we refer to $C(D)-R_m$ as the hiding redundancy,
which can reflect the hiding capability of the steganographic
code.

\subsection{Unicity Distance of Stego-key}

According to the Kerckhoff's principle, the security of a
steganographic code should be based on nothing but the secrecy of
the stego key. Therefore, it is important to analyze the key
equivocation. In details, we want to know how many data the
attacker must used to recover the stego key, i.e. the unicity
distance of stego key. We analyze this problem according to two
kinds of attacking conditions. One is stego-only extracting
attack, i.e. the attacker can only get the stego objects; the
other is known-cover extracting attack that means that the
attacker can get not only the stego objects but also some
corresponding cover objects. And we begin the analysis with
known-cover attack.

\bigskip
\begin{theorem}\label{thm3}
$({\cal M},f_N ,\phi _N )$ is length-$N$ steganographic code
subject to distortion $D$ with zero probability of error, i.e. for
any given $\varepsilon  > 0$, $P_{eN}  = P(\phi _N (X^N ,K) \ne M)
\le \varepsilon$. Then for given sequence of $n$ ($n$ is large
enough) pairs of cover objects and stego objects, the expectation
of spurious stego keys $\overline S _n$ for known-cover extracting
attack has the lower bound such that
\[\overline S _n  \ge \frac{{2^{H(K)} }}{{2^{nN(C(D) - R_m
+ \varepsilon )} }}-1 \enspace ,
\]
where $C(D)=\mathop{\max}\limits_{q(x|\widetilde{x}) \in Q}
H(X|\widetilde{X})$ is the hiding capacity and
$R_m=\frac{H(M)}{N}$ is the message rate.

\medskip
\begin{proof}
For a given sequence of pairs of cover objects and stego objects
$(\widetilde{x}^N ,x^N )^n $, the set of possible stego keys is
defined as
\begin{eqnarray*}
K((\widetilde{x}^N ,x^N )^n )=\{ k \in {\cal K}|\exists \,m^n
\in {\cal M}^n \;{\mbox{such that}}\\
P(m^n ) > 0\;{\mbox{and}}\,\,f_N^n (\widetilde{x}^{Nn} ,m^n ,k) =
x^{Nn} \}
\end{eqnarray*}
where
\begin{eqnarray*}
&&f_N^n(\widetilde{x}^{Nn} ,m^n ,k)\\
&=&(f_N (\widetilde{x}_1^N ,m_1 ,k), \cdots ,f_N
(\widetilde{x}_n^N ,m_n ,k))\\
&=&(x_1^N , \cdots ,x_n^N )=x^{Nn}
\end{eqnarray*}
So the number of spurious stego keys for observed
$(\widetilde{x}^N ,x^N )^n$ is $ \left|K((\widetilde{x}^N ,x^N )^n
)\right| - 1 $, and the expectation of spurious stego keys is
given by
\begin{eqnarray*}
\overline S _n  &=& \sum\limits_{(\widetilde{x}^N ,x^N )^n}
{P((\widetilde{x}^N ,x^N
)^n )\left[\left|K((\widetilde{x}^N ,x^N )^n )\right| - 1\right]}\\
&=& \sum\limits_{(\widetilde{x}^N ,x^N )^n} {P((\widetilde{x}^N
,x^N )^n )\left|K((\widetilde{x}^N ,x^N )^n )\right|} - 1 \enspace
.
\end{eqnarray*}
Using Jesen's inequality, we can get
\begin{eqnarray}
&&H(K|\widetilde{X}^{Nn} ,X^{Nn} ) \nonumber\\
&=&\sum\limits_{(\widetilde{x}^N ,x^N )^n } {P((\widetilde{x}^N
,x^N )^n )H(K|} \;(\widetilde{x}^N
,x^N )^n) \nonumber \\
&\le&\sum\limits_{(\widetilde{x}^N ,x^N )^n } {P((\widetilde{x}^N
,x^N )^n )\log _2 \left| {K((\widetilde{x}^N ,x^N )^n )}
\right|} \nonumber \\
&\le& \log _2 \sum\limits_{(\widetilde{x}^N ,x^N )^n }
{P((\widetilde{x}^N ,x^N )^n )\left| {K((\widetilde{x}^N ,x^N )^n
)}\right|}\nonumber\\
&=&\log _2 (\overline S _n  + 1) \enspace . \label{eqn2}
\end{eqnarray}
On the other hand, $f_N^n (\widetilde{x}^{Nn} ,m^n ,k) = x^{Nn}$
implies $H(X^{Nn} |\widetilde{X}^{Nn} ,M^n ,K) = 0$, which,
together with the assumption that key is independent of message
and cover object, message is independent of cover object, and the
sequences $\widetilde{X}^{Nn}$ and $M^n $ are both i.i.d. sequence
of random variables, yields that
\begin{eqnarray}
&&H(\widetilde{X}^{Nn} ,X^{Nn} ,M^n ,K) \nonumber\\
&=&H(X^{Nn}|\widetilde{X}^{Nn},M^n ,K) + H(\widetilde{X}^{Nn} ,M^n ,K)\nonumber\\
&=&H(\widetilde{X}^{Nn} ,M^n ) + H(K) \nonumber\\
&=&NnH(\widetilde{X}) + nH(M) + H(K)\enspace . \label{eqn3}
\end{eqnarray}
Since the steganographic code satisfies zero probability of error,
we have, for any given $\varepsilon  > 0$,
\begin{eqnarray}
&&P(\phi _N^n (X^{Nn} ,K) \ne M^n )\nonumber\\
&=& P((\phi _N (X_1^N,K), \cdots
,\phi _N (X_n^N ,K) \ne (M_1 , \cdots M_n ))\nonumber \\
&=&P(\exists \,i \mbox{ such that }1 \le i
\le n \mbox{ and } \phi _N (X_i^N ,K) \ne M_i)\nonumber \\
&\le&\sum\limits_{i = 1}^n {P(\phi_N (X_i^N ,K) \ne M_i )}\nonumber \\
&\le& n\varepsilon \enspace . \label{eqn4}
\end{eqnarray}
Equation (\ref{eqn4}) with Fano's inequality implies that for any
given $\varepsilon > 0$,
\begin{equation}\label{eqn5}
H(M^n |X^{Nn} ,K) \le n\varepsilon
\end{equation}
Furthermore, because sequence $\widetilde{X}^{Nn}$ is i.i.d.
sequence of random variables and cover channel is memoryless, we
obtain that
\begin{eqnarray}
&&H(\widetilde{X}^{Nn} ,X^{Nn} ,M^n ,K)\nonumber\\
&=&H(\widetilde{X}^{Nn} ) + H(X^{Nn} |\widetilde{X}^{Nn} ) +
H(K|\widetilde{X}^{Nn} ,X^{Nn} )\nonumber \\
&&{+} H(M^n |\widetilde{X}^{Nn},X^{Nn},K) \nonumber \\
&\le& N n H(\widetilde{X}) + N n H(X|\widetilde{X}) +
H(K|\widetilde{X}^{Nn}, X^{Nn} ) \nonumber \\
&&+ H(M^n |X^{Nn} ,K)\nonumber\\
&\le& N n H(\widetilde{X}) + N n H(X|\widetilde{X}) +
H(K|\widetilde{X}^{Nn} ,X^{Nn} ) + n\varepsilon . \label{eqn6}
\end{eqnarray}
combining (\ref{eqn3}) and (\ref{eqn6}) yields that, for any given
$\varepsilon > 0$,
\begin{equation}\label{eqn7}
H(K|\widetilde{X}^{Nn} ,X^{Nn} ) \ge H(K) + nH(M) -
NnH(X|\widetilde{X}) - n\varepsilon \enspace ,
\end{equation}
which, together with (\ref{eqn2}), implies for any given
$\varepsilon  > 0$,
\begin{displaymath}
\log _2 (\overline S _n  +1) \ge H(K) + nH(M) -
NnH(X|\widetilde{X}) - n\varepsilon \enspace ,
\end{displaymath}
i.e.
\begin{equation}\label{eqn8}
\overline S _n  \ge \frac{{2^{H(K)} }}{{2^{n(NH(X|\widetilde{X}) -
H(M) + \varepsilon )} }} - 1 \enspace .
\end{equation}
Since hiding capacity $C(D)$ satisfies $C(D) = \mathop {\max}
\limits_{q(x|\widetilde{x}) \in Q} H(X|\widetilde{X})$ and $R_m =
\frac{H(M)}{N}$, we have, for any given $\varepsilon  > 0$,
\begin{displaymath}
\overline S _n  \ge \frac{{2^{H(K)} }}{{2^{nN(C(D) - R_m +
\varepsilon )} }} - 1 \enspace .
\end{displaymath}
\end{proof}
\end{theorem}

\bigskip
\begin{definition}\label{def4}
The unicity distance $n_0$ for a steganographic code with
known-cover extracting attackers is the minimum number of pairs of
cover objects and stego objects with which one expects that the
expectation of spurious stego keys equals zero. And the unicity
distance $n_1$ for a steganographic code with stego-only
extracting attackers is the minimum number of stego objects with
which one expects that the expectation of spurious stego keys
equals zero.
\end{definition}
\bigskip

It is easy to know that $n_1  \ge n_0 $ . And using Theorem
\ref{thm3}, we can get the following important corollary.

\bigskip
\begin{corollary}\label{clr1}
The unicity distance $n_0$ for known-cover extracting attack and
$n_1$ for stego-only extracting attack satisfy that for any given
$\varepsilon > 0$,
\begin{displaymath}
n_1  \ge n_0  \ge \frac{{R_k }}{{C(D) - R_m  + \varepsilon }}
\enspace ,
\end{displaymath}
where $C(D) = \mathop {\max }\limits_{q(x|\widetilde{x}) \in Q}
H(X|\widetilde{X})$ is the hiding capacity, $R_m=\frac{H(M)}{N}$
is the message rate and $R_k = \frac{H(K)}{N}$ is the key rate.
\end{corollary}
\bigskip

Corollary \ref{clr1} shows that larger key rate $R_k$ and smaller
hiding redundancy $C(D) - R_m$ can make stronger difficulty of
extraction. The former is clear, while, for the latter, we give an
intuitive explanation as follows. Smaller hiding redundancy means
a message rate more appropriate for the cover channel. In this
case, dealing with the stego-objects (such as sampling) with
correct and spurious key respectively can only bring small
differences. In other words, it is difficult for the extracting
attacker to distinguish between the correct key and spurious ones.

\subsection{The Analysis for LSB Steganography}

As an example, we use the results in preceding subsection to
analyze the most popular steganographic mechanism, i.e. random LSB
steganoraphy on images, such as F5 \cite{wes}, Outguess \cite{pro}
and ``Hide and Seek'' \cite{mor}.

LSB replacing steganography usually work in the following manner:
Firstly, select an image with $N$ DCT coefficients for JPEG images
(or $N$ pixels for spatial images) denoted by
$C=(c_1,\cdots,c_N)$. Then randomly pick a subset of pixels, $\{
c_{j_1 }, \cdots ,c_{j_L } \}$, using a Pseudo-Random Number
Generator (PRNG) which is seeded with a stego-key $k$ belonging to
the key space $\mathcal{K}$, i.e. the PRNG with $k$ generates a
embedding path $\{ j_1,\cdots, j_L \}$. Finally, embedding the
message sequence $M=(m_1,\cdots,m_L)$, where $m_i \in \{ 0,1\}$,
by replacing the LSBs of $\{ c_{j_1 }, \cdots ,c_{j_L } \}$ or
other embedding operations such as $\pm 1$ to the DCT coefficients
(or pixels), and generate the stego-image $S=(s_1,\cdots,s_N)$.
Two kinds of embedding operations are shown in Table I and Table
II respectively.

\begin{table}[htbp]
\caption{LSB replacing embedding operation} \label{table1}
\centering
\begin{tabular}
{c|c|c|c|c} \hline Sample value& \multicolumn{2}{|c|}{2$i$} &
\multicolumn{2}{|c}{2$i$+1}  \\
\hline Embedded message bit& 0& 1& 0&
1 \\
\hline Modified sample value& 2$i$& 2$i$+1& 2$i$&
2$i$+1 \\
\hline
\end{tabular}
\end{table}

\begin{table}[htbp]
\caption{$\pm 1$ embedding operation} \label{table2}
\centering
\begin{tabular}
{c|c|c|c|c} \hline Sample value& \multicolumn{2}{|c|}{2$i$} &
\multicolumn{2}{|c}{2$i$+1}  \\
\hline Embedded message bit& 0& 1& 0&
1 \\
\hline Modified sample value& 2$i$& 2$i$+1 or 2$i$-1& 2$i$ or
2$i$+2&
2$i$+1 \\
\hline
\end{tabular}
\end{table}

The embedding rate $r$ is defined as the ratio of the length of
message to that of image, i.e. $r = \frac{L}{N}$. which means that
the possibility of a DCT coefficient (or pixel) being selected to
carry one bit message is $r$, because the message is asked to
randomly scattered in the whole image. Since message sequence $M$
is usually cipher text, we assume that $M$ is uniformly
distributed and independent with $C$, therefore every pixel is
modified with probability $\frac{r}{2}$. In fact LSBs of images
are similar to noise data and then approximately is uniformly
distributed and independent with $M$, so the assumption of
modifying rate being $\frac{r}{2}$ is also reasonable for plain
text $M$.

When using Corollary  \ref{clr1}, we have to compute the hiding
capacity that is hard generally. However, if the cover-objects are
binary sequence satisfying distribution of
Bernoulli$(\frac{1}{2})$ and the distortion metric is Hamming
metric, hiding capacity is given in \cite{mou:wan}. The capacity
is
\begin{equation}\label{eqn9}
C(D)=\left\{\begin{array}{ll} H(D) \quad & \mbox{if  } 0 \le D \le
\frac{1}{2} \\ 1 \quad & \mbox{if  } D > \frac{1}{2} \end{array}
\right . \enspace ,
\end{equation}
where $H(D)= -D\log_2 D - (1 - D)\log _2 (1 - D)$.

To analyze the LSB steganography, for simple we take the LSBs of
the DCT coefficients (or pixels) as cover-objects, which satisfies
distribution of Bernoulli$(\frac{1}{2})$ approximatively. And when
the the embedding rate is $r$ ($0\le r \le 1$), message rate is
just $R_m  = \frac{L}{N} = r\,{\mbox{bits/sign}}$ (note that $R_m$
has a unit but embedding rate $r$ has not) and the Hamming
distortion is $\frac{r}{2}$. Therefore (\ref{eqn9}) implies the
hiding capacity is $H(\frac{r}{2})$, and the hiding redundancy is
$H(\frac{r}{2})-r$.

\begin{figure}
\centering
\includegraphics[width=2.5in]{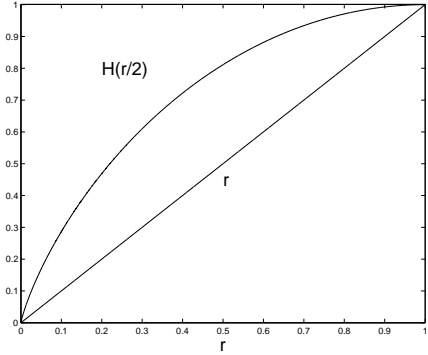}
\caption{Hiding redundancy: the curve denotes the ``hiding
capacity'' $H\left( {\frac{r}{2}} \right)$, the beeline stands for
the message rate $r$, and the difference between them is just the
hiding redundancy.} \label{fig:redundancy}
\end{figure}

In Fig.\ref{fig:redundancy}, it is clear that when $r \to 0$ (or
$r \to 1$), the redundancy of cover channel $H\left( {\frac{r}{2}}
\right) - r \to 0$, with which Corollary \ref{clr1} implies that
the unicity of the stego key tends to infinity, i.e. it is hard
for the attack to succeed.

\section{
Extracting Attack on LSB Replacing Steganography of Spatial
Images}

Reference \cite{fri:gol:sou} presents an extracting attack on LSB
steganography of JPEG images (such as F5 and Outguess), and
\cite{fri:gol:sou:hol} make an extracting attack on LSB (replacing
or $\pm 1$) steganography of spatial images. The purposes of these
attacks are both to recovery the stego-key, and the experimental
results show the same phenomena that the attacking processes need
more data for small or large embedding rate $r$, and when $r \to
0$ (or $r \to 1$) the attacks will fail, which consists with the
information-theoretic conclusion in Sect. II. However, on the
other hand, it should be noted that the analysis in Sect. II is
based on some general assumptions and the lower bound in corollary
\ref{clr1} is obtained from known-cover attack although it is also
a lower bound for stego-only attack. Therefore the results of
preceding section can only reflect the tendency of the difficulty
of recovering stego key but can not be used to estimate the amount
of needed data by the attacker. And the methods of
\cite{fri:gol:sou} and \cite{fri:gol:sou:hol} are both based on
non-parameter hypothesis testing, by which it is hard to calculate
the necessary amount of samples. Now we present a new stego key
searching method for LSB replacing steganography of spatial images
by using a parameter hypothesis testing, which is efficient and
simpler than preceding methods. The main contribution of our
attack is that it can accurately estimate the amount of necessary
data, which is important because with less data we cannot get the
stego key while too much data will slow down the searching speed.

Our method is also an example about how to do extracting attack by
combining traditional techniques of cryptanalysis and steganalysis
together. The main ideas are as follows. Firstly estimate the
length of the message (the embedding rate) with some detecting
methods. And then filter the stego image to get the data of its
noise area that can be thought of as a sample from a mixture
distribution \cite{wu} with the mixing parameter as a function of
the embedding rate. Through analyzing this mixture distribution,
we can exploit some ``accordant advantage" of the correct stego
key over those spurious ones. Finally, with this accordant
advantage, do the correlation attack as cryptanalysis to obtain
the stego key.

We do extracting attack under the assumption that we get a stego
image and know the steganographic algorithm. And the only thing we
don't know is just the stego key. This assumption is similar with
that in cryptanalysis. And in this paper, 8 bits grayscale images
is taken as examples to describe our method. And the same
notations as those in Sect. II (C) will be used. In details,
denote the cover image and stego image with $N$  pixels by
$C=(c_1,\cdots,c_N)$ and $S=(s_1,\cdots,s_N)$ respectively, where
$c_i,s_i \in [0,255]$ and $1 \le i \le N$. The stego key $k$,
belonging to the key space $\mathcal{K}$, is just the seed of the
PRNG. The message sequence is denoted by $M=(m_1,\cdots,m_L)$.
Notice that, as mentioned in Sect. I, message is usually required
to be encrypted before it is embedded into images, which is why
recovering stego key with simple brute-force search has to
consider the encryption key at the same time. And the purpose of
our method is to get the stego key $k$ regardless of encryption
key when getting only the stego image $S$.

\subsection{A mixture distribution model of stego images' noise}

LSB steganography essentially hides the message in the noise area
of the image. Therefore we analyze the noise data of the stego
image. Firstly filter the stego image $S=(s_1,\cdots,s_N)$ with
spatial average filter, and get a``new image" $\bar S = \{ \bar
s_1 ,\bar s_2 , \cdots ,\bar s_N \}$. Note that here save $\bar
s_i$'s as real numbers, i.e. keep several digits of decimal
fraction when averaging pixels. Then take difference between the
pixels of $S$ and $\bar S$ as the noise data. For $1 \le i \le N$,
if $s_i$ is odd, the noise data is defined as $w_i  = s_i  - \bar
s_i $, and if $s_i$ is even $w_i  = \bar s_i  - s_i $. The set of
noise data is denoted by $W = \{ w_1 ,\,w_2 , \cdots ,w_N \} $.

It is reasonable to assume that the noise data $w_i$'s
corresponding to $s_i$'s, which have not been modified, is a
sample from a Gaussian White Noise approximately, i.e. a normal
distribution with mean $0$ and variance $\sigma ^2$. And if the
pixel $s_i$ in $i$th position has been modified in embedding
process, $1$ has been added to $c_i$ when $s_i$ is odd, and $1$
has been subtract from $c_i$ when $s_i$ is even as shown in Table
\ref{table1}. Therefore $w_i$'s corresponding to modified $s_i$'s
can be viewed as a sample from a normal distribution with mean $1$
and the same variation $\sigma ^2$. Here we ignore the influence
of modifying pixels around the position $i$, because this kind of
influence is counteracted by averaging them. Both of the two
assumptions have been verified by experimental results on many
images. When embedding rate is $r$ , in $S$ on average
$\frac{r}{2}$ of pixels have been modified. So $W = \{ w_1 ,\,w_2
, \cdots ,w_N \} $ is a sample from a mixture distribution
\begin{equation}\label{eqn10}
F_{{\textstyle{r \over 2}}} (x) = (1 - \frac{r}{2})F(x) +
\frac{r}{2}G(x)
\end{equation}
where $F(x)$ and $G(x)$ are the distribution functions of normal
distribution $N(0,\sigma ^2 )$ and $N(1,\sigma ^2)$ respectively.

For $k \in \mathcal{K}$, let $I(k)$ denote the set of sample
indices visited along the path generated from the key $k$. If $k$
is a spurious key, $\{ w_j \} _{j \in I(k)} $ is a random sample
from distribution (\ref{eqn10}). On the other hand, if $k$ is just
the correct key $k_0$, in $\{ w_j \} _{j \in I(k_0)}$ on average
$50\%$ of samples are from distribution $F(x)$ and the other
$50\%$ of them from the distribution $G(x)$. So in this case, $\{
w_j \} _{j \in I(k_0)}$ is a random sample from mixture
distribution such as
\begin{equation}\label{eqn11}
F_{{\textstyle{1 \over 2}}} (x) = \frac{1}{2}F(x) +
\frac{1}{2}G(x)\enspace.
\end{equation}
When $0 < r < 1$, the difference between distributions
(\ref{eqn10}) and (\ref{eqn11}) can be used to distinguish the
correct key from those spurious ones.

\subsection{Accordant Advantage}

To exploit the difference between mixture distributions
(\ref{eqn10}) and (\ref{eqn11}), let $X_0$ be a random variable
with distribution function $F(x)$, $X_1$ is a random variable with
distribution function $G(x)$, $\alpha _0  = P\{ X_0  > A\} $, and
$\alpha _1  = P\{ X_1  > A\} $, where $A$ is a real number larger
than zero. Then
\begin{equation}\label{eqn12}
\alpha _0  = \int_A^{ + \infty } {dF(x)}  = \int_A^{ + \infty }
{\frac{1}{{\sqrt {2\pi } \sigma }}\exp\left\{ - \frac{{x^2
}}{{2\sigma ^2 }}\right\} dx}\enspace,
\end{equation}
\begin{equation}\label{eqn13}
\alpha _1  = \int_A^{ + \infty } {dG(x)}  = \int_A^{ + \infty }
{\frac{1}{{\sqrt {2\pi } \sigma }}\exp\left\{ - \frac{{(x - 1)^2
}}{{2\sigma ^2 }}\right\} dx}\enspace.
\end{equation}
Write $\Delta \alpha  = \alpha _1  - \alpha _0 $. It is easy to be
proved that $\Delta \alpha  > 0$.

As mentioned above, for the correct key $k_0$, the sample of noise
data set $\{ w_j \} _{j \in I(k_0)} $ can be modeled as the
realizations of a random variable $Y_0$ whose distribution
function is (\ref{eqn11}), while for an incorrect key $k$, sample
$\{ w_j \} _{j \in I(k)} $ can be viewed as the realizations of a
random variable $Y_1$  whose distribution function is
(\ref{eqn10}). Let $p_0  = P(Y_0  > A)$ and $p_1  = P(Y_1  > A)$,
then
\begin{equation}\label{eqn14}
p_0  = \int_A^{ + \infty } {dF_{{\textstyle{1 \over 2}}} (x)}  =
\frac{1}{2}\alpha _0  + \frac{1}{2}\alpha _1 \enspace ,
\end{equation}
\begin{equation}\label{eqn15}
p_1  = \int_A^{ + \infty } {dF_{{\textstyle{r \over 2}}} (x)}  =
(1 - \frac{r}{2})\alpha _0  + \frac{r}{2}\alpha _1 \enspace.
\end{equation}
And then the difference between them is that
\begin{equation}\label{eqn16}
\Delta p = p_0  - p_1  = \frac{1}{2}(1 - r)(\alpha _1  - \alpha _0
) = \frac{1}{2}(1 - r)\Delta \alpha \enspace .
\end{equation}
When the embedding rate $r$ being less than $1$, $\Delta p > 0$
because $\Delta \alpha  > 0$. That implies the correct key can
sample large noise data with lager possibility than a spurious key
does. Call $\Delta p$ as the ``accordant advantage". When $\Delta
p$ being large enough, we can recover the correct key. Given the
$r$, $\Delta p$ is determined by $\Delta \alpha$, therefore we
hope to take the proper $A$ to get the largest $\Delta \alpha$.
Define function
\begin{equation}\label{eqn17}
Q(x) = \frac{1}{{\sqrt {2\pi } }}\int_x^{ + \infty } {\exp\left\{
- \frac{{y^2 }}{2}\right\} dy} \enspace .
\end{equation}
Then $ \alpha _0  = Q({\textstyle{A \over \sigma }})$, $ \alpha _1
= Q({\textstyle{A-1 \over \sigma }})$, therefore $\Delta \alpha =
Q({\textstyle{{A - 1} \over \sigma }}) - Q({\textstyle{A \over
\sigma }})$. And when ${\textstyle{{A - 1} \over \sigma }} =  -
{\textstyle{A \over \sigma }}$, i.e. $A=\frac{1}{2}$, $\Delta
\alpha$ is largest. In this case,
\begin{equation}\label{eqn18}
\Delta \alpha  = Q( - {\textstyle{1 \over {2\sigma }}}) -
Q({\textstyle{1 \over {2\sigma }}}) = 1 - 2Q({\textstyle{1 \over
{2\sigma }}})\enspace .
\end{equation}

To compute the values of $p_0$ and $p_1$, we need also estimate
the variation $\sigma^2$. Denote the second moment of sample $W$
as $\bar a_2 $, i.e. $\bar a_2  = {\textstyle{1 \over
N}}\sum\limits_{i = 1}^N {w_i^2 }$. Notice that $W$ is the sample
from distribution (\ref{eqn10}), therefore the result in \cite{wu}
implies that $ \bar a _2  = (1 -
\frac{r}{2})(\mathord{\buildrel{\lower3pt\hbox{$\scriptscriptstyle\frown$}}
\over \sigma } ^2 + 0^2 ) +
\frac{r}{2}(\mathord{\buildrel{\lower3pt\hbox{$\scriptscriptstyle\frown$}}
\over \sigma } ^2 + 1^2 )$, i.e.
\begin{equation}\label{eqn19}
\mathord{\buildrel{\lower3pt\hbox{$\scriptscriptstyle\frown$}}
\over \sigma } ^2  = \bar a_2  - \frac{r}{2} \enspace.
\end{equation}
And we take statistic (\ref{eqn19}) as the estimation of
$\sigma^2$.

\subsection{Correlation Attack}

In this section, we borrow the idea of correlation attack in
cryptanalysis to recover the stego key with the accordant
advantage $\Delta p$. For $k \in \mathcal{K}$ the set of indices
generated from the key $k$ is denoted as $I(k) = \{ j_1 ,j_2 ,
\cdots ,j_L \}$. And the corresponding sample from noise set $W$
obtained with $k$ is $\{ w_{j_1 } ,w_{j_2 } , \cdots ,w_{j_L } \}
$ which can be viewed as a sequence of i.i.d. (independent and
identically distributed) random variables. Define a new sequence
of random variables as
\begin{displaymath}
Z_i  = \left\{ {\begin{array}{*{20}c}
   {1,\quad {\rm{if}}\;w_{j_i }  > A}  \\
   {0,\quad {\rm{if}}\;w_{j_i }  \le A}  \\
\end{array}} \right.,\quad 1 \le i \le L \enspace .
\end{displaymath}
Therefore $Z_i$'s are also i.i.d random variables. Construct a
sequence of statistics such as $\eta _n  = \sum\limits_{i = 1}^n
{Z_i } $ where $1 \le n \le L$. For the correct key $k_0$, the
analysis in Sect. III (B) shows that $P\{ Z_i  = 1\}  = p_0 $ ,
and the Central Limit Theorem implies that the distribution of
$\eta _n$ is approximately equal to the normal distribution
$N(np_0 ,np_0 (1 - p_0 ))$ when $n$ is large enough. Similarly, on
the other hand, for an incorrect key $k$, the distribution of
$\eta _n$ is approximately equal to normal distribution $N(np_1
,np_1 (1 - p_1 ))$ when $n$ is large enough. Then the work of
searching the correct key can be formulated as the following
hypothesis testing problem:
\medskip
\begin{itemize}[\setlabelwidth{$H_0$:}]
\item[$H_0$:] $\eta _n \sim N(np_0 ,np_0 (1 - p_0 ))$ which means
$k$ is just the correct key $k_0$;
\item[$H_1$:] $\eta _n \sim
N(np_1 ,np_1 (1 - p_1 ))$ which means $k$ is an incorrect key.
\end{itemize}
Select a threshold $T$. If $\eta _n  \ge T$, accept $H_0$,
otherwise accept $H_1$.
\medskip

Generally larger number of samples $n$ we use, more accurate
decision we can do. However, larger $n$ means spending more
searching time. We should determine $n$ and the threshold $T$ so
as to achieve the proper probability of the false alarm event
$p_f$ and that of missing event $p_m$. Using (\ref{eqn17}), we
obtain that
\begin{equation}\label{eqn20}
p_f  = Q\left( {\frac{{T - np_1 }}{{\sqrt {np_1 (1 - p_1 )} }}}
\right),\quad p_m  = Q\left( {\frac{{np_0  - T}}{{\sqrt {np_0 (1 -
p_0 )} }}} \right)
\end{equation}
In the present problem, we mainly concern $p_f$. When the number
of all possible stego keys is $|\mathcal{K}|$, $p_f$ is picked as
small as ${\frac{1}{2^{|{\cal K}|}}}$ so that the correct key  can
be determined uniquely. And $p_m$ could be chosen close to zero
(for example $10^{-2}$). For given $p_f$ and $p_m$, search the
Table for Standard Normal Distribution Function to get $w_f$ and
$w_m$ such that ${\textstyle{1 \over {2^{|{\cal K}|} }}} = Q(w_f
)$ and $p_m  = Q(w_m )$. Then with (\ref{eqn20}), we can compute
the needed values of $n$ and $T$ as follows:
\begin{equation}\label{eqn21}
n = \left[ {\frac{{w_m \sqrt {p_0 (1 - p_0 )}  + w_f \sqrt {p_1 (1
- p_1 )} }}{{\Delta p}}} \right]^2 \enspace ,
\end{equation}
\begin{equation}\label{eqn22}
T = w_f \sqrt {np_1 (1 - p_1 )}  + np_1 \enspace .
\end{equation}
Note that to get $n$ samples of noise data, $n^\ast$
($n^\ast\approx \frac{n}{r}$) pixels are needed on average. So
combining (\ref{eqn16}) and (\ref{eqn21}), we can get an
estimation for the number of needed pixels $n^\ast$ such as
\begin{equation}\label{eqn23}
n^*  \approx \frac{{4\left( {w_m \sqrt {p_0 (1 - p_0 )}  + w_f
\sqrt {p_1 (1 - p_1 )} } \right)^2 }}{{r[(1 - r)\Delta \alpha ]^2
}} \enspace .
\end{equation}
Equation (\ref{eqn23}) shows that $n^\ast\rightarrow\infty$ as
$r\rightarrow 0 \mbox{ or } 1$. In other words, when the embedding
rate $r$ is very small (close to $0$) or very large (close to
$1$), the process of recovering stego key will become difficult
because we have not enough pixels to use. Notice that this is
accordant with the information-theoretic analysis in Sect.II. And
this conclusion will also be proved by he experimental results on
``Hide and Seek 4.1" in next section.

With preparations above, now we describe the attacking method.
Assume that we have detect a stego image $S$ with $N$ pixels, and
know details of the steganographic algorithm except the stego key.
The attacking procedure goes through the following steps.

\bigskip
\emph{\textbf{Algorithm -- Correlation Attack}}
\begin{itemize}[\setlabelwidth{Step 4}] \item[Step 0]
\begin{enumerate}[\setlabelwidth{4)}]
\item Estimate embedded message length $L$ and the embedding rate
$r$ ($r=\frac{L}{N}$) using the method in \cite{jf:mg}; \item
Filter the stego image $S$ and take the noise data set $W = \{ w_1
,\,,w_2 ,\, \cdots ,w_N \} $ as described in Sect. III (A); \item
Estimate the variance $\sigma^2$ with statistic (\ref{eqn19}). Let
$A=0.5$, and compute $p_0$ and $p_1$ by using equations
(\ref{eqn12}), (\ref{eqn13}), (\ref{eqn14}) and (\ref{eqn15});
\item Let $p_f  = {\textstyle{\frac{1}{2^{|{\cal K}|} }}}$, choose
a proper $p_m$ (for example $10^{-2}$), and pick the $w_f$ and
$w_m$ such that ${\textstyle{\frac{1}{2^{|{\cal K}|} }}} = Q(w_f
)$, and $p_m  = Q(w_m )$. Finally compute the necessary number of
samples $n$ and the threshold $T$ using (\ref{eqn21}) and
(\ref{eqn22}).
\end{enumerate}

\item[Step 1] If $n>L$ , go to Step 3; otherwise, test all stego
keys in $\mathcal{K}$: for every $k \in \mathcal{K}$, seed the
PRNG with $k$ to generate the set containing $n$ sample indices
$I(k) = \{ j_1 ,j_2 , \cdots ,j_n \}$ and extract $n$ samples of
noise data $\{ w_{j_1 } ,w_{j_2 } , \cdots ,w_{j_n } \} $. Then
count the number $T_k$ of $w_{j_i }$'s such that $w_{j_i }
> 0.5$, i.e. $T_k  = |\{ w_{j_i } |\,w_{j_i }  > 0.5,\,1 \le i \le n\}
|$. If $T_k  < T$, reject $k$, otherwise save $k$ to the set $B$,
i.e. $B = \{ k|\,k \in {\cal K}\,{\mbox{ and }}\,T_k  \ge T\} $.

\item[Step 2] If $|B| = 1$, then then take the only key in $B$ as
the correct key and stop; If $|B|=0$ or $|B|>1$ go to Step 3;

\item[Step 3] Let $n=L$. Test all keys in $\mathcal{K}$ as does in
step $1$ and obtain $T_k $ for every $k \in {\cal K}$. Write
$T_{\max }  = \mathop {\max }\limits_{k \in {\cal K}} \{ T_k \} $,
and $D = \{ k|\,k \in {\cal K}\,{\mbox{ and }}\,T_k  = T_{\max }
\}$;

\item[Step 4] If $|D|=1$, then take the only key in $D$ as the
correct key and stop; If $|D|>1$, the attack fails and stop.
\end{itemize}

\section{Extracting Attack on ``Hide and Seek 4.1"}

As an example, we use our method to recover the stego key of "Hide
and Seek 4.1" \cite{mor} which is a typical LSB replacing
steganographic algorithm on the GIF file with 256 shades of gray
or color (In fact the deviser of ``Hide and Seek" suggest that
greyscale is best by far). The PRNG, used in ``Hide and Seek" to
generate the embedding path, is based on the function ``random (
)" of ``Borland C++3.1", which is seeded by a seed of 16 bits and
the length of message together. Hiding program encrypts the header
information, which consists of the 16 bits seed, length of message
and number of version, with IDEA cipher to produce 64 bits cipher
texts and embeds them into the LSBs of the first 64 pixels of the
GIF file. The key of IDEA is generated by a password consisting of
not more than 8 characters (64 bits). Therefore the receiver, who
knows the password, can decipher the hider information to get the
seed and length of message, which will seed the PRNG to extract
the hidden message.

It is hard to recover the 64 bits key of IDEA, but we can skip the
first 64 pixels and recover the key of PRNG with ``Correlation
Attack'' directly. ``Hide and Seek 4.1" uses only GIF images with
$320\times 480$ pixels, so the maximum length of message is
defined as 19000 bytes.\footnote{ In ``Hide and Seek",when used as
a part of key, the unit of message's length is byte.} And the
approach of \cite{jf:mg} can estimate the embedding rate with
error between $\pm 0.02$, therefore mostly about 760 ($19000\times
0.04$) possible lengths need to be tested when searching for the
key. In other words, the cardinality of the key space we search is
$2^{16}\times760$, i.e. the length of virtual key is only about 26
bits ($16+ \log_2 760 \approx 25.57$).

We do the experiment on 40 GIF files with 256-greyscale for
several kinds of embedding rates. And the correct key can be
determined when embedding rate $r$ satisfies $5.3\%  < r < 94.7\%
$. However, because the image used by ``Hide and Seek" is small
(only $320\times480$ pixels), for $|\mathcal K| = 2^{16}  \times
760$, the number of needed samples $n$ usually is larger than $L$,
the algorithm has to do the Step 3. To test the estimations for
$n$ and $T$ with (\ref{eqn21}) and (\ref{eqn22}), we also do the
experiment under the assumption that the length of message being
known, which means the key is only the 16 bits of seed. In this
case, for $r$ such that $1.1\% < r < 98.4\% $, we can get the
correct key successfully. Plain text and cipher text are embedded
respectively with ``Hide and Seek 4.1'' for the experiments and
the attacking results are similar. These Experiments are achieved
on Pentium IV machines running at 2.4GHz, 512MB RAM, and there is
a search rate of 250-8400 keys per second. The search speed is
greatly influenced by the embedding rate.

The detailed results of experiments on lena.gif and peppers.gif,
when key is only the 16 bits of seed, list Table \ref{table3} and
Table \ref{table4} respectively. In the tables,  ``-" means that
estimated number of samples $n$ is larger than the length of
message $L$, and the attack will do Step 3; $T_{k_0}$ with ``*" is
smaller than threshold $T$ and $|B|$ is zero, therefore the attack
also will do the Step 3. It is shown that, when $r$ satisfying
$10.5\% < r < 52.6\%$ (i.e. $200 \le L \le 9000$), the necessary
number of samples $n$ is smaller than the length of message $L$,
and the attacking procedure can stop successfully in step 2. In
this case, there is searching speed increase of $10\% - 45\% $
than that of setting $n=L$ directly, and note that the $T_{k_0}$
is larger than but close to the threshold $T$, which implies that
the necessary number of samples $n$ and the threshold $T$ obtained
with (\ref{eqn21}) and (\ref{eqn22}) are accurate.

And on the whole the attacking processes need more data for
smaller or larger embedding rate $r$, and when $r\rightarrow 0$
(or $r \rightarrow 1$) attacks will fail, which verifies the
information-theoretic conclusion in Sect. II once more.

\begin{figure}[h]
\begin{minipage}{0.25\textwidth}
\centerfigcaptionstrue \centering
\includegraphics[scale=0.2]{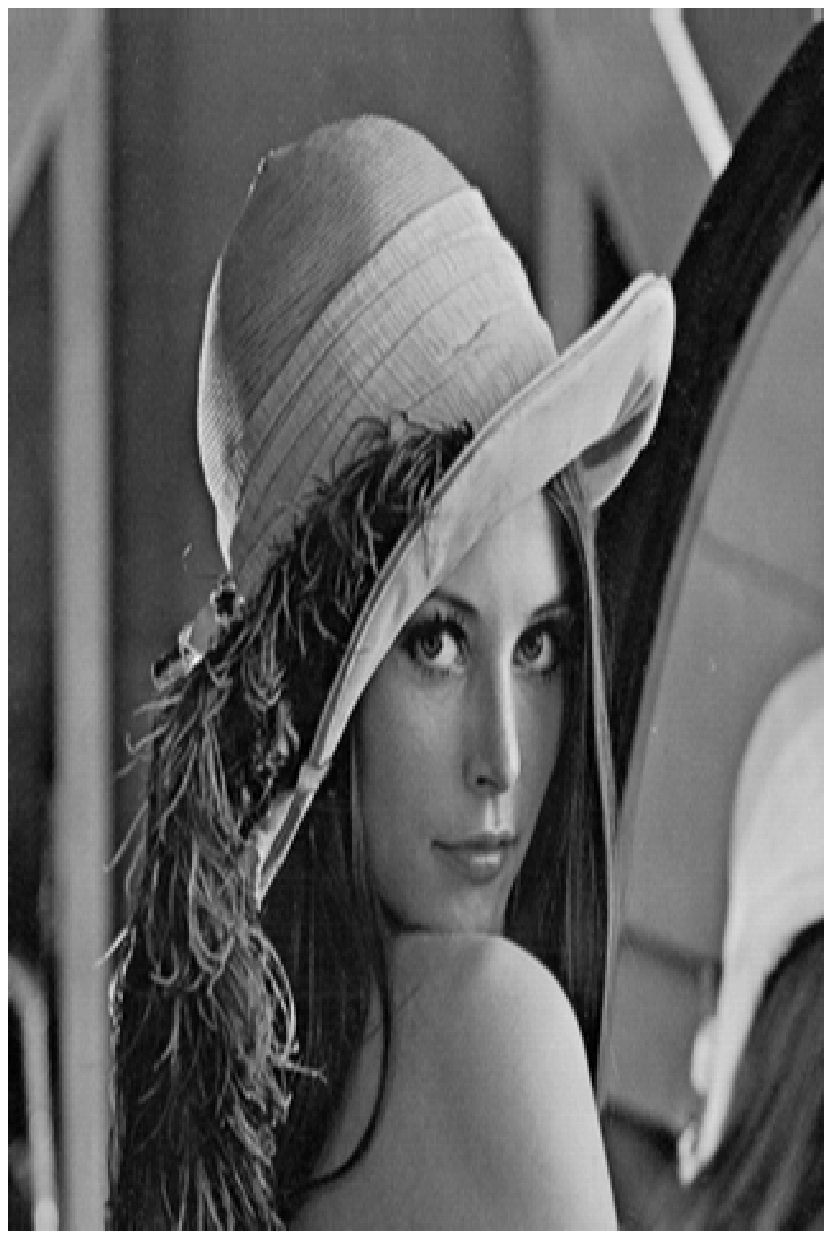}\\*
\caption{lena.gif}%
\end{minipage}
\begin{minipage}{0.25\textwidth}
\centerfigcaptionstrue \centering
\includegraphics[scale=0.2]{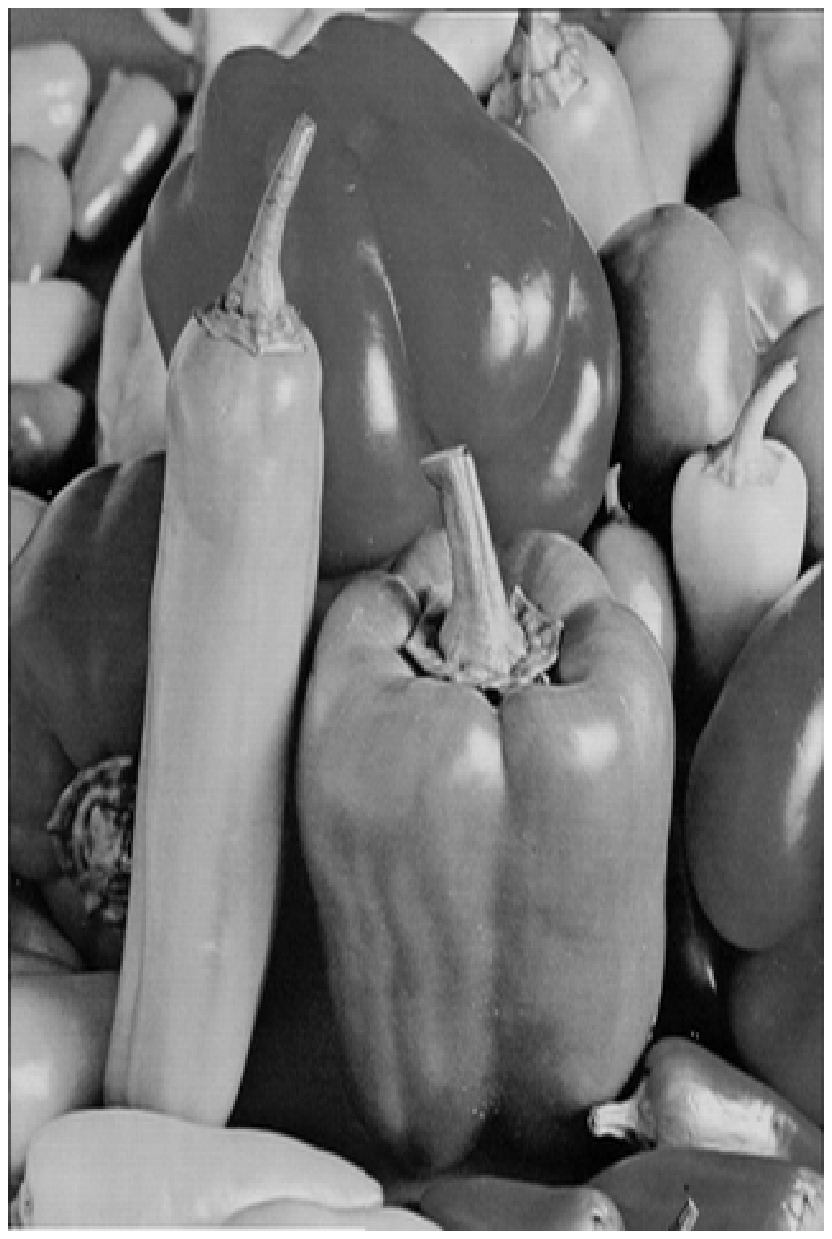}\\*
\caption{peppers.gif}
\end{minipage}
\end{figure}

\begin{table*}[htbp]
\renewcommand{\arraystretch}{1.3}
\caption{Experimental results on lena.gif} \label{table3}
\centering
\begin{tabular}{|c|c|c|c|c|c|}
\hline Length of & Embedding rate & Number of  & Threshold T &
$T_{k_0}$ corresponding  &
Result of attack \\
 message L (bytes)& $r$ & samples n (bytes)& & to the correct key $k_0$&\\
\hline 100 & 0.005 & -- & -- & -- & Fail\\
\hline 200 & 0.011 & -- & -- & -- & Succeed\\
\hline 1000 & 0.053 & -- & -- & -- & Succeed\\
\hline 2000 & 0.105 & 1830 & 6925 & 7086 & Succeed\\
\hline 3000 & 0.158 & 2066 & 7845 & 8040 & Succeed\\
\hline 5000 & 0.263 & 2699 & 10319 & 10466 & Succeed\\
\hline 8000 & 0.421 & 4374 & 16888 & 16922 & Succeed\\
\hline 9000 & 0.474 & 5293 & 20503 & 20560 & Succeed\\
\hline 10000 & 0.526 & 6535 & 24086 & 25398 & Succeed\\
\hline 12000 & 0.632 & 10805 & 41954 & 42265 & Succeed\\
\hline 13000 & 0.684 & -- & -- & -- & Succeed\\
\hline 18700 & 0.984 & -- & -- & -- & Succeed\\
\hline 18800 & 0.989 & -- & -- & -- & Fail\\
 \hline
\end{tabular}
\end{table*}

\begin{table*}[htbp]
\renewcommand{\arraystretch}{1.3}
\caption{Experimental results on peppers.gif} \label{table4}
\centering
\begin{tabular}{|c|c|c|c|c|c|}
\hline Length of & Embedding rate & Number of  & Threshold T &
$T_{k_0}$ corresponding  &
Result of attack \\
 message L (bytes)& $r$ & samples n (bytes)& & to the correct key $k_0$&\\
\hline 50 & 0.003 & --& -- & -- & Fail\\
\hline 100 & 0.005 & -- & -- & -- & Succeed\\
\hline 200 & 0.011 & -- & -- & -- & Succeed\\
\hline 1000 & 0.053 & -- & -- & -- & Succeed\\
\hline 2000 & 0.105 & 1301 & 4874 & 5008 & Succeed\\
\hline 3000 & 0.158 & 1470 & 5527 & 5669 & Succeed\\
\hline 5000 & 0.263 & 1921 & 7341 & 7982 & Succeed\\
\hline 8000 & 0.421 & 3111 & 11894 & 11933 & Succeed\\
\hline 9000 & 0.474 & 3764 & 14368 & 14493 & Succeed\\
\hline 10000 & 0.526 & 4648 & 17889 & 17964 & Succeed\\
\hline 12000 & 0.632 & 7680 & 29912 & 29514* & Succeed\\
\hline 13000 & 0.684 & -- & -- & -- & Succeed\\
\hline 18700 & 0.984 & -- & -- & -- & Succeed\\
\hline 18800 & 0.989 & -- & -- & -- & Fail\\
 \hline
\end{tabular}
\end{table*}

\section{Conclusion}

In the field of steganalysis, so far there have been many
literatures about detecting attack while there are few about
extracting attack. But the latter also will be concerned greatly
because it is a problem that a cryptanalyst has to face. In this
paper, we make a preliminary analysis on this problem using
information-theoretic method that is an analogue of Shannon's for
cryptography \cite{sha}. And the results can give some general
idea about the extracting attack no steganogrphy.

Our basic idea is that the extracting attack is in principle a
kind of cryptanalysis, and it should rely on both steganalysis and
cryptanalysis. As an example, we present an effective extracting
method no popular LSB replacing steganography of spatial images by
using the detecting technique of steganalysis and correlation
attacking technique of cryptanalysis together. The analysis for
our extracting method and the experimental results on ``Hide and
Seek 4.1" are both accordant with the information-theoretic
conclusion.

Better lower bounds on unicity of stego key for stgeo-only attack
and attacks under other conditions are interesting problems that
we will study. And our further work will also include exploiting
extracting approaches on other kinds of steganographic algorithms.

\appendices



\section*{Acknowledgment}
This paper is supported by NSF of China No 60473022. And the
authors would like to thank Jia Cao, Ning Ma, Wei Guan and Heli
Xiao for many helpful and interesting discussions.




\begin{thebibliography}{1}
%
\bibitem{cac}
C. Cachin, ``An information-theoretic model for steganography,''
in Information Hiding: Second International Workshop, vol. 1525 of
LNCS, Springer-Verlag, 1998, pp. 306--318.
%
\bibitem{zol:fed}
J. Z\"{o}llner, H. Federrath, H. Klimant, A. Pfitzmann, R.
Piotraschke, A. Westfeld, G. Wicke, and G. Wolf, ``Modeling the
security of steganographic systems,'' in Information Hiding, 2nd
International Workshop, vol. 1525 of LNCS, Springer-Verlag, 1998,
pp.344-354.
%
\bibitem{kat:pet}
S. Katzenbeisser and F. A. Petitcolas, ``On defining security in
steganographic systems,'' Security and Watermarking of Multimedia
Contents IV, vol. 4675. Proceedings of SPIE, 2002, pp. 260-268.
%
\bibitem{hop:lan}
N. J. Hopper, J. Langford, and L. van Ahn, ``Provably secure
steganography,'' in Advances in Cryptology: CRYPTO 2002, vol. 2442
of LNCS, Springer-Verlag, 2002, pp.18-22
%
\bibitem{ett}
M. Ettinger, ``Steganalysis and game euilibria,'' in Information
Hidinging, Second International Workshop, vol. 1525 of LNCS,
Springer-Verlag, 1998, pp. 319-328.
%
\bibitem{som:mer}
A. Somekh-Baruch, and N. Merhav, ``On the capacity game of public
watermarking systems,'' (2002) Available:
http://tiger.technion.ac.il/\symbol{126}merhav/papers/p71.ps
%
\bibitem{mou:sul}
P. Moulin, and J. A. O'Sullivan, ``Information theoretic analysis
of information hiding,'' IEEE Trans. on Information Theory, vol.
49, 2003, pp. 563-593.
%
\bibitem{vol:pun}
S. Voloshynovskiy, and T. Pun, ``Capacity-security analysis of
data hiding technologies,'' in IEEE International Conference on
Multimedia and Expo ICME2002, Lausanne, Switzerland, August, 2002.
pp. 26-29.
%
\bibitem{cha:mem}
R.Chandramouli, and N. D. Memon: Steganography capacity: A
steganalysis perspective. Proc. of SPIE on Security and
Watermarking of Multimedia Contents V, Vol. 5020, 2003, pp.
173-177.
%
\bibitem{mos:cha:new}
I. S. Moskowitz, L. Chang,and R. E. Newman, ``Capacity is the
wrong paradigm'' (2002) Available:
http://chacs.nrl.navy.mil/publications/CHACS/2002/2002moskowitz-capacity.pdf
%
\bibitem{zha:li}
W. M. Zhang, and S. Q. Li, ``Security measurements of
steganographic systems,'' in The Second International Conference
of Applied Cryptogarphy and Network Security, Vol. 3089 of LNCS,
Springer-Verlag, Berlin Heidelberg New York, 2004, pp. 194-204.
%
\bibitem{wes:pfi}
A. Westfeld, and A. Pfitzmann, ``Attacks on Steganographic
Systems,'' In 3rd International Workshop. vol. 1768 of LNCS,
Springer-Verlag, Berlin Heidelberg New York, 2000, pp. 61-75
%
\bibitem{fri:gol}
J. Fridrich, M. Goljan and R. Du, ``Attacking the outguess,''
Proceedings of the ACM Workshop on Multimedia and Security,
France, 2002, pp. 3-6
%
\bibitem{zha:pin}
T.Zhang, and X. J. Ping, ``A new approach to reliable detection of
LSB steganography in natural images,'' Signal Processing, Elsevier
Science, Vol.83, No.10, 2003, pp. 2085-2093
%
\bibitem{cha}
R. Chandramouli, ``A mathematical framework for active
steganalysis,'' In ACM Multimedia Systems Journal, Special Issue
on Multimedia Watermarking, ACM Multimedia Systems Journal,
Special Issue on Multimedia Watermarking, vol. 9, no. 3, 2003, pp.
301-311,
%
\bibitem{fri:gol:sou}
J. Fridrich, M. Goljan, and D. Soukal, ``Searching for the stego
key,'' Security, Steganography and Watermaking of Multimedia
Contents of EI SPIE, Vol. 5306, 2004, pp. 70-82.
%
\bibitem{wes}
A. Westfeld, ``High Capacity Despite Better Steganalysis (F5-A
Steganographic Algorithm),'' In: LNCS, vol. 2137, Springer-Verlag,
New York, pp. 2001, 289-302.
%
\bibitem{pro}
N. Provos, ``Defending Against Statistical Steganalysis,'' 10th
USENIX Security Symposium, Washington, DC, (2001), Available:
http://www.stanford.edu/class/ee380/Abstracts/011107.html
%
\bibitem{fri:gol:sou:hol}
J. Fridrich, M. Goljan, D. Soukal, and T. Holotyak, `` Forensic
Steganalysis: Determining the Stego Key in Spatial Domain
Steganography,'' Proc. EI SPIE San Jose, CA, 2005, pp. 631-642 .
%
\bibitem{ma:zha}
N. Ma, W. M. Zhang, W. F. Liu, `` Extracting attack to LSB
stganography of JPEG images,''  Proc. Of the Tenth Joint
International Computer Conference. International Academic
Publishers, World Publishing Corporation. China, 2004, 336-340.
%
\bibitem{sie}
T. Siegenthaler, ``Decrypting a class of stream ciphers using
ciphertext only,'' IEEE Transactions on Computers, vol.C-34, 1985,
81-85.
%
\bibitem{mor}
Shaggy, ``Hide and Seek,'' (2005) Available:
http://www.jjtc.com/Security/stegtools.htm
%
\bibitem{spy}
Spyware Information Center Report, (2005) Available:
http://www3.ca.com/securityadvisor/pest/pest.aspx?id=2601
%
\bibitem{mou:wan}
P. Moulin, Y. Wang, ``New results on steganographic capacity,"
Proceeding of CISS 2004. University of Princeton, Princeton, New
Jersey (2004) Availabel:
http://www.ifp.uiuc.edu/\~ywang11/paper/CISS04\_204.pdf
%
\bibitem{wu}
W.R. Wu, ``Estimation of Parameters in A Mixture of Two Normal
Distributions," Journal of Fujian Agricultural College, Fujian,
China, Vol.18, No. 2, 1989, pp. 236-243.
%
\bibitem{jf:mg}
J. Fridrich and M. Goljan, ``On Estimation of Secret Message
Length in LSB Steganography in Spatial Domain," Proc. EI SPIE ,
Vol. 5306, Security, Steganography, and Watermaking of Multimedia
Contents VI, 2004, pp. 23-34.

%
\bibitem{sha}
C. E. Shannon, ``Communication theory of secrecy system," Bell
Syst. Tech. J., vol. 28, 1949, pp. 656-715.

\end{thebibliography}
%

\begin{biography}{Weiming Zhang}
was born in Hebei, P. R. China in 1976. He is working for the Ph.D
degree in Cryptology in Zhengzhou Information Engineering
University. His research interests include probability theory,
cryptology, and information hiding.
\end{biography}

\begin{biography}{Shiqu Li}
 was born in Sichuan, P. R. China in 1945. He received his
MSc. degrees in probability theory from Beijing Normal University,
P. R. China in 1981. He is currently a Professor in the Department
of Applied Mathematics at Zhengzhou Information Engineering
University. His primary research interests include probability
theory, cryptology, and especially the logic funtions in
cryptology.
\end{biography}



\vfill


\end{document}